\documentstyle[twocolumn,prl,aps]{revtex}

\begin{document}

\twocolumn[\hsize\textwidth\columnwidth\hsize
 \csname @twocolumnfalse\endcsname

\title{Chiral non-linear $\sigma$-models as models for topological
superconductivity}

\author{A.G.~Abanov$^1$ and P.B.~Wiegmann$^{2,3}$\\ }

\address{
    ${}^1$~{\it Department of Physics,
    Massachusetts Institute of Technology,
    77 Massachusetts Ave.,
    Cambridge, MA 02139 }\\
    ${}^2$~{\it James Franck Institute
    of the University of Chicago,
    5640 S. Ellis Ave.,
    Chicago, IL 60637 }\\
    ${}^3$~{\it
    Landau Institute for Theoretical Physics}
}


\maketitle

\begin{abstract}
We study the mechanism of topological superconductivity in a
hierarchical chain of chiral non-linear $\sigma$-models (models of
current algebra) in one, two, and three spatial dimensions. The models
have roots in the 1D Peierls-Fr\"{o}hlich model and illustrate how the
1D Fr\"{o}hlich's ideal conductivity extends to a genuine
superconductivity in dimensions higher than one.  The mechanism is
based on the fact that a point-like topological soliton carries an
electric charge.  We discuss a  flux quantization mechanism and
show that it is essentially a generalization of the persistent current
phenomenon, known in quantum wires.  We also discuss why the
superconducting state is stable in the presence of a weak disorder.
\end{abstract}

\pacs{PACS number(s): }
]
\vfill


{\it Introduction.} In this paper we discuss models of topological
superconductivity, where fermionic currents are dynamically coupled
with topological textures of chiral bosonic fields \cite{topsup}.
The superconducting state in these models emerges due to an unbroken
chiral symmetry.  Electronic off-diagonal correlations in topological
superconductors are different from those of conventional BCS-type
superconductors, but rather close to those of Luttinger liquids
\cite{WTS3}.  (i) Topological superconductivity takes place in
standard models of current algebras (\ref{2+1}-\ref{3+1}) (also called
chiral non-linear $\sigma$-models), and  (ii) the hydrodynamic
action (and wave functions of low energy states) possesses a geometric
phase (or $\vartheta$-term) -- an imaginary part of the Euclidian
action
\begin{equation}
  \label{g}
     {\cal S}_{\rm E}
   =  2\pi\lambda_L^2\int dx\,dt\, j_\mu^2
    +i\pi N_f{\vartheta}.
\end{equation}
Here $j_\mu=(v_0\rho,\vec j)$ is an electronic current and density
deviation from the equilibrium (we set $v_0=1$ in the following),
$\lambda_L$ is the London's penetration depth, and $N_f$ is a
degeneracy of an electronic state (e.g., spin).

The first term in (\ref{g}) gives the London equations.  $\vartheta$
is an integer valued topological invariant. The theta term of
(\ref{g}) does not affect the equations of motion but the monodromy of
the wave function. It is the geometric phase which carries an
information about spacetime transformation properties of excitations.
In particular, it changes by $ 1$ in the processes of a
$2\pi$-rotation of a group of $N_f$ electrons around some axis or an
interchange of positions of two groups of $N_f$ electrons. In a
multi-connected sample, it also changes by $ 1$ under the translation
of $N_f$ electrons along a noncontractable path. We show that this
term provides a mechanism for flux quantization in a multi-connected
geometry.  The geometric phase is, perhaps, the most important
difference from the s-wave BCS theory, where the Euclidian action is
real.  It emphasizes a quantum interference aspect of the
mechanism. While discussing the $\vartheta$-term, we rely on the
results of Ref.\cite{AbanovWiegmann-1999}.


{\it  Models of current algebras.} The models
of topological superconductivity in two and three spatial
dimensions are the standard non-linear $\sigma$-models of current
algebras (see e.g.
\cite{currentalgebras}).  These
models describe Dirac fermions interacting with a chiral boson field
\begin{eqnarray}
   \label{2+1}
      {\cal L}_2 &=& \bar{\psi}\left(i\hat D
        -m \vec{n}\cdot\vec{\tau}\right)\psi+W_2[\vec n],\\
  \label{3+1}
      {\cal L}_3 &=& \bar{\psi}\left[ i\hat D
        -m (\pi_0+i\gamma_5\vec\pi\cdot\vec{\tau})
	\right]\psi+W_3[\pi],
\end{eqnarray}
where $i\hat D=\gamma_{\mu}(i\partial_\mu +A_\mu)$ and $\gamma_\mu$
are $2\times 2$ and $4\times 4$ Dirac matrices in spatial dimensions
two and three.  The fermions are $SU(2)$ doublets and
$\vec\tau=(\tau_1,\,\tau_2,\,\tau_3)$ are the Pauli matrices.  In
addition, a fermion variable $\psi$ carries a (suppressed) flavor
index running from 1 to $N_f$.  The parity of the number of flavors
$N_f$ plays an important role in what follows.  The chiral fields take
values on two and three-dimensional spheres thus matching the number
of spatial dimensions: $\vec n^2=1$ and $\pi_0^2+\vec\pi^2=1$.  We
test the system by an external electromagnetic potential $A_\mu$.
Finally, $W_{2,\,3}$ describe a bare action of the chiral field, other
than the dynamics induced by fermions. In 2D, e.g., it is $
W_2=\frac{M}{8\pi}({\dot{\vec n}}^2- s^2(\partial_j\vec n)^2)+\cdots
$. The models have an electronic origin and can be viewed as an effective
description of certain interacting electronic systems. We do not
discuss that here.

We emphasize that the essential properties of these models  
are inherited from the model of Fr\"ohlich's ideal conductivity
in one spatial dimension \cite{Frohlich-1954}. We, therefore, add the
latter to the hierarchy.  In this model, 1D fermions interact with a
chiral field taking values on a circle:
\begin{equation}
 \label{1+1}
      {\cal L}_1 =\bar{\psi}\left(i\hat D
        -m e^{i\gamma_5\phi}\right)\psi.
\end{equation}
It is well known \cite{Frohlich-1954} that $2\pi$-phase solitons of
the Fr\"ohlich model are charged and  move freely through the
system, making it an ideal conductor.  It is not a superconductor,
however, since the restricted spatial geometry does not allow for a
transverse motion of a charge flow.  As a result, a weak disorder or
commensurability effects destroy the ideal conductivity\cite{ALR}.  We
will see that in dimensions greater than one (models
(\ref{2+1},\ref{3+1})) a
phenomenon similar to Fr\"ohlich's ideal conductivity results in 
superconductivity even in the presence of  weak disorder.

The physics of the models (\ref{2+1}-\ref{1+1}) is controlled by the chiral
symmetry. In 1D it is: $\phi\to\phi+\alpha,\;\;\psi\to
e^{-\frac{i}{2}\gamma_5\alpha}\psi$.   In 2D it is: $\vec n\vec\tau\to
U^{-1 }
\vec n\vec\tau U$ and $\psi\to U^{-1}\psi$, where $U$ is an $SU(2)$
matrix.  In 3D it is $g\to U^{-1} g U$ and $\psi\to
U^{-\gamma_5}\psi$, where $g=\pi_0+i\vec\pi\vec\tau$.

We argue that if the global chiral symmetry is {\em unbroken}, the ground
states of the 2D and 3D models are superconducting.  These models then
exhibit a Meissner effect and a flux quantization with period
$\frac{\Phi_0}{N_f}$ if $N_f$ is even and $\frac{\Phi_0}{2N_f}$ if $N_f$
is odd, where $\Phi_0=hc/e$ is the flux quantum.  Notice that the period
is always a fraction of $\Phi_0$ with an even denominator regardless of
whether the number of fermionic species is even or odd.  For the most
interesting cases of polarized ($N_f=1$) and unpolarized ($N_f=2$)
electrons, the flux is quantized with the period $\Phi_0/2$.


{\it Electronic and topological currents.} In every model
(\ref{2+1}-\ref{1+1}) a coordinate space matches the target space of
chiral fields.  Therefore, there exist spatial configurations of a
chiral field which wrap over the target space with an arbitrary integer
winding number $Q$ -- topological charge.  They are solitons: phase
solitons in 1D and
skyrmions in $D=2,3$.  The topological charge $Q$ may
be written as a spatial integral $Q=\int d^Dx\,J_0$ of the zeroth
component of topological current $J_\mu$:
\begin{eqnarray}
  \label{curr1}
     J_\mu &=& \frac{1}{2\pi} \epsilon_{\mu\nu}\partial_\nu\phi,
  \\
  \label{curr2}
     J_\mu &=& \frac{1}{8\pi} \epsilon_{\mu\nu\lambda}
      \vec n\cdot\partial_\nu\vec n\times\partial_\lambda\vec n,
  \\
  \label{curr3}
     J_\mu &=&
     \frac{1}{12\pi^2}\epsilon_{\mu\nu\lambda\rho}
     \epsilon_{abcd}\pi_a\partial_\nu\pi_b
        \partial_\lambda\pi_c\partial_\rho\pi_d .
\end{eqnarray}%
In configurations without singularities, a topological current
conserves $\partial_\mu J_\mu=0$.  This reflects the fact
that topological charge cannot be changed by smooth variations of
chiral fields.

It is known that the energy spectra of the models
(\ref{2+1}-\ref{1+1}) in a smooth static soliton background with a
topological charge $Q$ have an additional $N_fQ$ levels with negative
energy, compared to the spectrum in a topologically trivial
background.  Being filled, these levels create an electric charge
localized on the solitons.  The motion of the solitons then results in an
electric current since the electric charge moves together with a
soliton as long as the motion is adiabatic \cite{fermionicurrent}.
This is summarized by
\begin{equation}
 \label{12}
     \int j_0 d^Dx=N_fQ;\;\;\;j_\mu(x)=N_fJ_\mu(x)+\cdots,
\end{equation}
where $j_\mu=\langle\bar\psi\gamma_\mu\psi\rangle$ is an electronic
current in a background chiral field and $J_\mu$ is a topological
current of a  chiral field (\ref{curr1}-\ref{curr3}).  The density of
the electronic current  (\ref{12}) has corrections (suppressed by the
gap $m$) due to terms with higher gradients.  However, the corrections do
not have a topological nature.  The change of total charge in a soliton
background is exact.


{\it Spin and momentum of a soliton and the $\vartheta$-term.}
Intuitively, it is appealing that a soliton carrying a fermionic charge
$N_fQ$ has a spin $N_fQ/2$ and statistics that depend on whether $N_fQ$
is even or odd.  It has been shown in
Refs.\cite{Jaroszewicz-1987,AbanovWiegmann-1999} that this is indeed
true -- an intrinsic quantum number of a soliton -- a fermionic number
(\ref{12}) converts into rotational and translational quantum numbers.
Spin and statistics of solitons are described by a geometric phase of
the wave function.  The geometric phase changes by $\pi N_fQ$ under an
adiabatic $2\pi$-rotation of a soliton or an adiabatic exchange of
the positions of two solitons.  In a multi-connected geometry, the
geometric phase also changes by the same amount under an adiabatic
translation of a soliton along a noncontractable cycle. Along such a
path,
the wave function acquires a phase $e^{i\pi N_fQ}$, which indicates
the spin and statistics of a soliton.  The geometric phase also results in
a shift in quantization of the orbital momentum $L=N_f\frac{Q}{2}+l$
and momentum $\Pi_i=\frac{2\pi}{L_i}(N_f\frac{Q}{2}+n_i)$ along a
noncontractable cycle (here $l$ and $n_i$ are integers and $L_i$ is
the length of the cycle).

We write fermionic parts of our models as ${\cal L}=\bar\psi{\cal
D}\psi$ with ${\cal D}$ an operator, dependent on chiral fields.  Then
the induced action ${\cal W}$ of the non-linear $\sigma$-model is a
result of an integration over fermions in a fixed background
configuration of a chiral field $e^{-{\cal W}}=\int e^{-\int{\cal
L}dxdt}d\psi d\bar\psi=(\mbox{Det}{\cal D})^{N_f}$. Then the geometric
phase appears as an imaginary part of the Euclidian action of a
non-linear $\sigma$-model. Indeed, in
Refs.\cite{Jaroszewicz-1987,AbanovWiegmann-1999} it has been shown
that the geometric phase in the models (\ref{2+1}-\ref{1+1}) is equal
to 
$N_f\pi$ times a topological invariant $\vartheta$ that characterizes
a topology of spacetime configurations of the chiral fields $\phi,\;\vec
n,\;(\pi_0,\vec\pi)$.  The $\vartheta$-term is also referred to as a
global chiral anomaly, distinct from the local anomaly given by (\ref{12}).

The $\vartheta$-term depends on the compactification of the
Euclidian spacetime.  If it is  a sphere, the $\vartheta$-invariant is a
representation of the homotopy groups $\displaystyle{\pi_{D+1}(S^D)}$.  It
is zero in 1D, any integer (the Hopf number) in 2D, and 0 or 1 in 3D.

In a more physical case, the Euclidian time is a circle and the
coordinate space may also have some noncontractable cycles.  The
topological $\vartheta$-term then is corrected by terms describing the
motion of solitons along the cycles. The explicit expression for the
$\vartheta$-term requires a higher profile of the topology
\cite{Pontrjagin-1941three}, but its physical meaning is simple. E.g.,
an adiabatic rotation of a soliton by an angle $\phi$ results in the
$\vartheta$-term $\vartheta [\phi]=Q\int\dot{\phi}\frac{dt}{2\pi}$
(see \cite{AbanovWiegmann-1999} and references therein). If a soliton
is adiabatically translated along a noncontractable spatial cycle, the
$\vartheta$-term is $\vartheta [\phi]=Q\int\dot{ u}\frac{dt}{2\pi}$,
where $u$ is a displacement of the soliton. The latter is the only
term which remains in 1D, where it is known to be responsible for the
parity effect in the persistent current phenomenon observed in quantum
wires \cite{perscurrent}.


{\it Non-linear $\sigma$-models.} Fermionic number (\ref{12}) and the
$\vartheta$-term determine the imaginary part of the Euclidian action
of a non-linear $\sigma$-model.  Its real part is given by a regular
gradient expansion. To leading order in $1/m$, we have (see e.g.,
\cite{AbanovWiegmann-1999} and references therein)
\begin{eqnarray}
 \label{nl1}
    \frac{{\cal W}_1}{N_f}&=&i\pi  \vartheta[\phi]+\int d^2x
     \left(iA_\mu J_\mu
    + \frac{1}{8\pi}(\partial_\mu\phi )^2\right),\\
 \label{nl2}
    \frac{{\cal W}_2}{N_f}
     &=& i\pi \vartheta[\vec n]+\int d^3x \left(iA_\mu J_\mu
     +\frac{m}{8\pi}\,(\partial_\mu\vec n)^2\right),\\
 \label{nl3}
    \frac{{\cal W}_3}{N_f}
    &=& i\pi  \vartheta[g]
    + \int d^4x \left(iA_\mu J_\mu+\frac{F_\pi^2}{2}
    {\rm tr}\,(\partial_\mu g^{-1}\partial_\mu g)\right),
\end{eqnarray}
where $F_\pi^2=\frac{1}{2\pi^2}m^2\ln\frac{\Lambda}{m}$ with $\Lambda$
being an ultraviolet cutoff and $g=\pi_0+i\vec\pi\vec\tau$  a
unitary matrix.  The first two terms represent global and local chiral
anomalies.


{\it London's hydrodynamics.} The hydrodynamics of a super\-conductor
is defined by a rigidity with respect to transverse charge currents
$\langle j_\mu (k),j_\nu
(-k)\rangle=(4\pi\lambda_L^2e^2)^{-1}(\delta_{\mu\nu}-\frac{k_\mu
k_\nu}{k^2})$ at $k\to 0$ and can be summarized by the London
Lagrangian: ${\cal L}=2\pi\lambda_L^2 j_\mu^2$.

In topological liquids, charge current is equal (at small $k$) to the
topological current (\ref{12}), so that a criterion for
superconductivity translates to a transverse rigidity of topological
currents
\begin{equation}
 \label{g1}
    \langle J_\mu (k),J_\nu (-k)\rangle
    =(4\pi N_f^2\lambda_L^2e^2)^{-1}\left(\delta_{\mu\nu}
    -\frac{k_\mu k_\nu}{k^2}\right).
\end{equation}

In 1D the first terms of the gradient expansion (\ref{nl1}) already give
(\ref{g1}).  We will argue that the transverse rigidity of topological
currents takes place in any dimension if the global chiral symmetry is
unbroken, i.e., if the chiral fields are in a quantum disordered state:
$\langle\vec n(x),\vec n(0)\rangle\to 0$ at large $x$ in 2D, or
$\langle g(x),g^{-1}(0)\rangle\to 0$ in 3D (it is always true in 1D
model).

The situation in higher dimensions is obscured by the fact that the
models (\ref{2+1}-\ref{3+1}) are not renormalizable and that the symmetry
of the ground state depends on the effect of small distances not
included in the continuum effective model.  To implement the state
with unbroken chiral symmetry (a quantum disordered state), one might
choose a non-electronic part $W_{2,3}$ of the models
(\ref{2+1}-\ref{3+1}).  The latter does not contribute to the
anomalous terms. In addition, we assume that the correlation radius of
the quantum disordered state $r_c$ is large so that the chiral field
is smooth and the electronic density follows the topological density
according to eq.~(\ref{12}).

The arguments that a quantum disordered state of a non-linear
$\sigma$-model, regardless of its dimension, is rigid with respect to
topological excitations, are general\cite{Polyakov}.  Here we adopt
them for the 2D model following Ref.\cite{Schakel91}.  We first write
the non-linear $\sigma$-model (\ref{nl2}) in the ${\rm CP}^1$
representation by writing $(\partial_\mu\vec n)^2$ as
$4z^\dagger(-i\partial_\mu-a_\mu)^2z$, where $z=(z_1,z_2)$ is a
complex doublet with constraint $z^\dagger z=1$ related to $\vec n$ as
$\vec n=z^\dagger \vec \tau z$.  The field strength of the gauge field
$a_\mu$ is proportional to the topological current of $\vec n$
according to $J_\mu=\frac{1}{2\pi}\epsilon^{\mu\nu\lambda}
\partial_\nu a_\lambda$ -- it describes the fluctuations of the solid
angle formed by the $\vec n$-field. The rigidity of topological currents
(\ref{g1}) means that the propagator of the gauge field $a_{\mu}$
induced by $z$-bosons is massless.

It is obvious that if the chiral field $\vec n$ is ordered, than
$z$-bosons are condensed and $\langle J_\mu (k),J_\nu
(-k)\rangle\sim\kappa k^2
$, where the stiffness $\kappa$ is proportional to the amount of the
condensate of z-bosons.  This combined with (\ref{12}) leads to an
insulating behavior.

On the contrary, if the chiral symmetry is unbroken the system does not
respond to the constant transverse field $a_\mu$. This means that 
the gauge field is massless $\langle a_\mu (k),a_\nu (-k)\rangle\sim
(r_c k^2)^{-1}$, which leads to (\ref{g1}) with $r_c
\sim\lambda_L^2e^2$\cite{Schakel91,suppression}.

Combining (\ref{g1}), the fermionic number of solitons, and the
geometric phase (\ref{nl1}-\ref{nl3}), we obtain the hydrodynamic
description of a topological superconductor (\ref{g}) in D=2,3 and
Fr\"ohlich's ideal conductor in D=1. In terms of topological currents,
it is
\begin{equation}
  \label{g2}
     {\cal S}
   = N_f\int (2\pi N_{f}\lambda_L^2 e^{2}J_\mu^2
   + ieJ_\mu A_\mu)\, dxdt+i\pi N_f{\vartheta}.
\end{equation}


{\it Disorder.} It is generally correct that if the Meissner effect
takes place in a system with no disorder, then a superconducting state
remains intact in the presence of  weak disorder, regardless of the
mechanism that led to superconductivity.  In 1D there are no
transverse directions and no true Meissner effect.  As a result the
Fr\"ohlich ideal conductivity is destroyed by the backscattering of a
single impurity \cite{ALR}.  A $2k_f$-harmonic of the impurity
potential $V_{2k_f}$ adds the term $|V_{2k_f}|\cos(\phi-\alpha)$ to
the effective Lagrangian (\ref{nl1}), where $\alpha$ is the phase of
the potential.  It pins a soft translational mode of the charge
density wave at $\phi=\alpha$, makes fluctuations of $\phi$ massive,
and suppresses correlations of topological currents at small momenta.

In higher dimensions, the existence of transverse currents eviscerates
this effect. In 2D, similar to the 1D case, we add a backscattering
term $\vec V_{2k_f}\vec n$ to (\ref{nl2}), where $\vec V_{2k_f}$ is a
vector, representing a $2k_f$ harmonic (scattering between Dirac
points of the spectrum) of an impurity potential. This term has a
tendency to align the $\vec n$-field along $\vec V_{2k_f}$, but does not
change the correlation of topological currents (solid angles of $\vec
n$) as long as the potential is smaller than a scale provided by the
correlation radius $r_c$. The system remains in the quantum disordered
phase, while the superfluid density is reduced by the amount of the
order $\vec V_{2k_f}^2$. One can see this formally by writing the
impurity term in ${\rm CP}^1$ representation $ \vec
V_{2k_f}z^\dagger\vec \tau z$. This term shifts the local chemical
potentials of the $z$ bosons in opposite directions for the bosons with
isospin along and opposite to $\vec V_{2k_f}$. If $z$-bosons are far
from the condensation point (ordering of $\vec n$), the small change of
chemical potential cannot drive the condensation, and fluctuations of
an internal gauge field $a_{\mu}$ remain massless. The superconducting
state cannot be destroyed by an infinitesimal amount of
impurities. The important difference with 1D is that the impurity
potential does not couple to a topological density of solitons
(i.e. to solid angle of $\vec n$-field in this case).  Similar
arguments are valid for the 3D model.


{\it Flux quantization.}  The geometric phase $i\pi N_f \vartheta$ in
the low energy action (\ref{g}) provides the mechanism of the flux
quantization in topological superconductors in a multi-connected
geometry. Consider a sample in the shape of a torus, threaded by magnetic
flux $\Phi$ through some particular torus cycle, which we label by
1. Due to the gauge invariance all energy levels are $\Phi_0$-periodic
functions of $\Phi$. We will show that the energy of a macroscopic
sample at a fixed chemical potential is periodic with a smaller
period: $\Phi_0/N_f$ if $N_f$ is even  and
$\Phi_0/2N_f$ if $N_f$ is odd.

Starting from the low energy action (\ref{g2}), we pass to the
Hamiltonian in a sector with a given topological charge $Q$.  We
introduce displacements $\vec u$ according to $J_0=-\vec\nabla\vec
u,\;\;\vec J=\dot{\vec u}$.  The relevant part of the $\vartheta$-term
can be written in terms of zero modes (the homogeneous part) of
displacements $\bar u_i=\int u_id^Dx$. E.g, a soliton moving around the
cycle 1 contributes $\vartheta=Q\int\frac{\dot{\bar
u}_1}{L_1}dt$ (similar to the one in 1D, where
$\bar{u}_1/L_1=\phi/2\pi$).

Neglecting all excitations transverse to the torus cycle pierced by
the flux, we set all currents except for $J_1$ to zero. Then we
neglect all harmonics of the displacement $u_1$ except for the zero
mode. Their contribution to the energy does not depend on the flux.
Setting the gauge potential to be a constant along the cycle 1, $\vec
A=(2\pi\Phi/L_1\Phi_0,\,0,\,0)$, we obtain the flux dependent part of
the Lagrangian (we return to the real time)
$$
    L=\frac{2\pi N_f^2\lambda_L^2e^2}{V}
    \left( {\dot{\bar u}_1}^2-Q^2\right) -2\pi N_f
    \left(\frac{Q}{2}+\frac{\Phi}{\Phi_0}\right)\frac{\dot{\bar u}_1}{L_1}.
$$
Varying it with respect to $\dot{\bar u}_1$ we find canonical momentum
$\Pi_1=4\pi N_f^2\lambda_L^2e^2 \frac{\dot{\bar u}_1}{V}
-\frac{2\pi N_f}{L_1}(\frac{Q}{2}+\frac{\Phi}{\Phi_0})$, where we observe
a shift of the momentum by the $\vartheta$-term.  Representing the
canonical momentum of the zero mode as
$\Pi_1=-i\frac{\partial}{\partial\dot{\bar u}_1}$, we obtain the wave
function of the zero mode $\exp{(i\frac{2\pi n_1}{L_1}\bar u_1)}$,
where $n_1$ is an integer,
and the flux dependent part of the energy of the sliding charge density
mode
\begin{equation}
 \label{e}
    E_{n_1}=\frac{\pi }{2\lambda_L^2e^2}
    \frac{V}{L_1^2}\left(\frac{n_1}{N_f}\!+\!\frac{\Phi}{\Phi_0}
    \!+\!\frac{Q}{2}\right)^{\!2}\!\!+\!2\pi N_f^2\lambda_L^2
    e^2\frac{Q^2}{V}.
\end{equation}
If $N_f$ is even, it is obvious that the energy is a $\Phi_0/N_f$-
periodic function of the flux (c.f., \cite{DorMav92}). However, the energy
remains $\Phi_0/2$-periodic even at $N_f=1$.  In fact, if $N_f$ is odd
the energy is $\Phi_0/2N_f$-periodic. In macroscopic samples, the energy
of vacuum polarization by a soliton -- the last term in (\ref{e}) -- is
negligible compared to the kinetic energy.  If $N_f$ is odd, the
ground state possesses no solitons $Q=0$ as long as
$\Phi<\Phi_0/4N_f$. If, however, $\Phi_0/4N_f<\Phi<\Phi_0/2N_f$, the
ground state at the fixed chemical potential acquires a topological
charge $Q=1$ and $N_f$ additional particles. The freedom to create a
soliton, when the flux exceeds $\Phi_0/4N_f$, halves the naive period
$\Phi_0/N_f$ of the kinetic energy.

In the 1D setup (a quantum wire) $V=L_1$, $\lambda_L^2e^2=1/4N_f$, and
(\ref{e}) gives the persistent current $j=-\partial E/\partial\Phi$
for the Peierls-Fr\"{o}hlich model (\ref{1+1}).  In this case the
vacuum polarization energy caused by a static soliton (the last term
in eq.(\ref{e})) destroys periodicity for odd $N_f$ (c.f.,
\cite{perscurrent}).

This mechanism is to be compared with the familiar flux quantization
in the BCS theory. In that case, a flux-dependent part of the energy is
given by zero modes of the phase $\bar\varphi=2\pi n \frac{x}{L_1}$ of
the order parameter along a noncontractable cycle: $E_n=\frac{\pi
V}{2\lambda_L^2 e^2 L^2}
\left(\frac{n}{2}+\frac{\Phi}{\Phi_0}\right)^2$. It is 
$\Phi_0/2$-periodic and has deep minima at fluxes which are multiples of
$\Phi_0/2$. Although the topological mechanism leads to a similar
result, the state counting and the underlying physical picture is
different.


We would like to thank R.B. Laughlin, P.A. Lee, A.M. Polyakov, and
X.-G. Wen for discussions.  A.A. was supported by NSF DMR
9813764. P.W. was supported by grants NSF DMR 9971332 and MRSEC NSF
DMR 9808595.


\end{document}